\documentclass[aps,pre,groupedaddress,twocolumn,10pt]{revtex4-1}
\usepackage{graphicx}
\usepackage{epstopdf}
\usepackage[english]{babel}
\usepackage[T1]{fontenc}

\usepackage{makecell}
\usepackage{color}
\usepackage{multirow}
\usepackage{slashbox}
\usepackage{booktabs}
\usepackage{amsmath}
\usepackage{amssymb}
\usepackage{amsthm}
\usepackage{bbm}
\usepackage{hyperref}
\usepackage[T1]{fontenc}
\usepackage[usenames,dvipsnames]{xcolor}
\hypersetup{colorlinks=true, linkcolor=BrickRed, urlcolor=blue!50!black, citecolor=blue!50!black}

\usepackage[normalem]{ulem}

\renewcommand{\imath}[0]{\mathsf{i}}

\usepackage[nodayofweek,level]{datetime}

\begin{document}

\title{Formation of a fast "lane" for positional transition in a microparticle-suspended nematic liquid crystal cell}

\author{Ke Xiao, Xi Chen, Xuezheng Cao and Chen-Xu Wu$^{*}$}

\affiliation{Department of Physics, School of Physical Science and Technology, Xiamen University, Xiamen 361005, People's Republic of China}
\email{cxwu@xmu.edu.cn}

\begin{abstract}
In this paper, based on the numerical calculation of total energy utilizing the Green's function method, we found that the external electric field applied to a microparticle-suspended nematic liquid crystal cell, if reaching a critical value, combined with its direction, surface anchoring feature and molecular dielectric anisotropy, is possible to create an anisotropic "bubble" around the microparticle with a vertical fast "lane", in which the microparticle can, driven by the asymmetric buoyant force,  vertically move swiftly from the cell's midplane to a new equilibrium position, triggering a positional transition discovered by the author previously. Such a new equilibrium position is decided via a competition between the buoyant force and the effective force built upon the microparticle by the elastic energy gradient along the "lane". The threshold value of external field, depends on thickness $L$ and Frank elastic constant $K$ and slightly on the microparticle size and density, in a Fr\'{e}edericksz-like manner, but by a factor. For a nematic liquid crystal cell with planar surface alignment, a bistable equilibrium structure for the transition is found when the direction of the applied electric field is (a) perpendicular to the two plates of the cell with positive molecular dielectric anisotropy, or (b) parallel to the two plates and the anchoring direction of the cell with negative molecular dielectric anisotropy. Except for the formation of a vertical fast "lane", when the electric field applied is parallel to both the two plates and perpendicular to the anchoring direction, the microparticle suspended in the nematic liquid crystal tends to be trapped in the midplane, regardless of the sign of the molecular dielectric anisotropy. Such phenomenon also occurs for negative molecular dielectric anisotropy while the external is applied perpendicular to the two plates. Explicit formulae proposed for the critical electric field agrees extremely well with the numerical calculation.

\end{abstract}
\date{\today}

\maketitle

\section{INTRODUCTION}
Liquid crystals (LCs) are soft matter with anisotropic properties characterized by their long-range orientational order.
In a nematic liquid crystal (NLC) phase, for example, molecules with molecular long axes possess a preferred orientation, which drives them to align along a common direction.
While in a dispersed NLC, colloidal particles disturb the alignment of LC molecules, inducing elastic distortions which give rise to long-range anisotropic interactions and topological defects. Over the past two decades, the properties and behaviors of colloids-suspended NLC have attracted considerable interest and triggered a wide range of promising practical applications, such as in new display and topological memory devices~\cite{B.Comisky1998,Z.Wang2011,T.Araki2011}, new materials~\cite{I.I.Smalyukh2018}, report external triggers and release microcargo~\cite{Kim2018}, and biological detectors~\cite{E.A.Nance2012,S.J.Woltman2007}.
To better understand the physics of colloidal particles embedded within an NLC cell, much effort has been made by means of experiment, theoretical modeling, and computer simulation so far~\cite{P.Poulin1997PRL,M.Vilfan2008,U.Ognysta2008,M.Skarabot2008,A.V.Ryzhkova2015,C.P.Lapointe2009,C.P.Lapointe2010,U.M.Ognysta2011,S.J.Kim2016,D.Andrienko2003,K.Izaki2013,T.Araki2017,C.Conklin2018}.

Experimentally, diverse methods and techniques have been developed to measure the interaction force between particles in NLC in a direct manner~\cite{P.Poulin1997PRL,F.L.Calderon1994,M.Yada2004,K.Takahashi2008,M.Skarabot2018}. Also, it has been found that the interaction force between spherical particles suspended in NLC is associated not only with interparticle distance and geological confinement~\cite{M.Vilfan2008}, but also with shape of particles which plays a crucial role in pair interaction and aggregation behaviors~\cite{C.P.Lapointe2009}. In the presence of an external electric field, additionally, fascinating physical phenomena such as levitation, lift, bidirectional motion, aggregation, electrokinetic and superdiffusion have been found for colloids dispersed in NLCs~\cite{O.P.Pishnyak2007,O.P.Pishnyak2011,C.Conklin2018,JM.Pages2019}.
In most cases the inclusion of particles into an NLC cell tends to create LC alignment singularities around the suspended substances, which in general are determined by boundary conditions such as surface anchoring features, particle size and shape, LC elasticity, and external fields etc~\cite{C.P.Lapointe2009,H.Stark2001,H.Stark2002,Y.Wang2017,X.Yao2018}.
It has been widely confirmed and accepted that when a spherical particle is immersed in NLC, there are three possible types of defect configurations~\cite{P.Poulin1997Since,P.Poulin1998,T.C.Lubensky1998}. Dipole and quadrupolar configurations are the two configurations usually seen around a spherical particle with strong homeotropic surface anchoring, whereas boojum defect is formed by a micro-sphere with homogeneous surface anchoring.
In addition, recently B. Senyuk and his coauthors assumed that conically degenerated boundary condition gives rise to the so-called elastic hexadecapole~\cite{B.Senyuk2016}, and reported that the dipole-hexadecapole transformation can be achieved via tuning the preferred tilt angle of LC molecules anchored on colloidal particle surface~\cite{Y.Zhou2019}.
Through experimental observations it has been also found that, when an external field is applied, there exists a transition between elastic dipole and quadrupolar configuration, which depends on particle size and surface anchoring strength~\cite{R.W.Ruhwandl1996,R.W.Ruhwandl1997,J.C.Loudet2001}.

On the other hand, theoretical modeling and computer simulation provide a useful complement to experimental investigations. The system of colloidal particles dispersed in NLCs are typically modelled via Landau-de Gennes (LdG) theory and elastic free energy method.
Furthermore, Monte Carlo simulation~\cite{T.Araki2017,N.Atzin2018}, lattice Boltzmann method~\cite{C.Denniston2001,S.Changizrezaei2019} and finite element method~\cite{M.Vilfan2008,M.Ravnik2009,M.Ravnik2011,M.Tasinkevych2012,S.R.Seyednejad2013} are commonly adopted techniques to minimize the LdG free energy functional. Except for the methods mentioned above, recently S. B. Chernyshuk and coauthors studied the interaction between colloidal particles in NLCs near one wall and in an NLC cell with or without an external field by using the method of Green's function, and obtained general formulae for interaction energy between colloidal particles~\cite{S.B.Chernyshuk2010,S.B.Chernyshuk2011,S.B.Chernyshuk2012}.
In the liquid crystal-particle coexistence system, it is found that an external field applied is able to drive particles apart~\cite{P.Poulin1997PRL}, rotation~\cite{C.P.Lapointe2010}, and alignment~\cite{G.DAdamo2015}, and even trigger a positional transition~\cite{K.Xiao2019}, which can be used to manipulate the suspended microparticle.
Although in an NLC cell the interaction of two particles is very well understood and the particle-wall interaction has been widely observed experimentally for a single particle immersed~\cite{S.B.Chernyshuk2011,S.J.Kim2014,B.K.Lee2017}, the properties of a single particle in such an NLC cell in the presence of an external electric field theoretically have not been fully addressed, partly due to the difficulties in mathematics involved in analyzing such kind of confined systems. However, it is of crucial importance to investigate, analytically if possible, the one microparticle-suspended NLC cell if one wants to find the applications of such kind of manipulations as found in experiments.

\section{THEORETICAL MODELING}
We begin our investigation by considering a system of a spherical microparticle of radius $r$ suspended in a NLC cell with $L$-thick spacers in the presence of an external electric field. For simplicity, the polarization of the microparticle is neglected compared to the influence of external field on the realignment of liquid crystal molecules. Fig. 1 schematically illustrates two systems under an external field respectively with (a) a homeotropic anchoring and (b) a homogeneous planar anchoring at the two cell walls.
\begin{figure}[htp]
  \includegraphics[width=\linewidth,keepaspectratio]{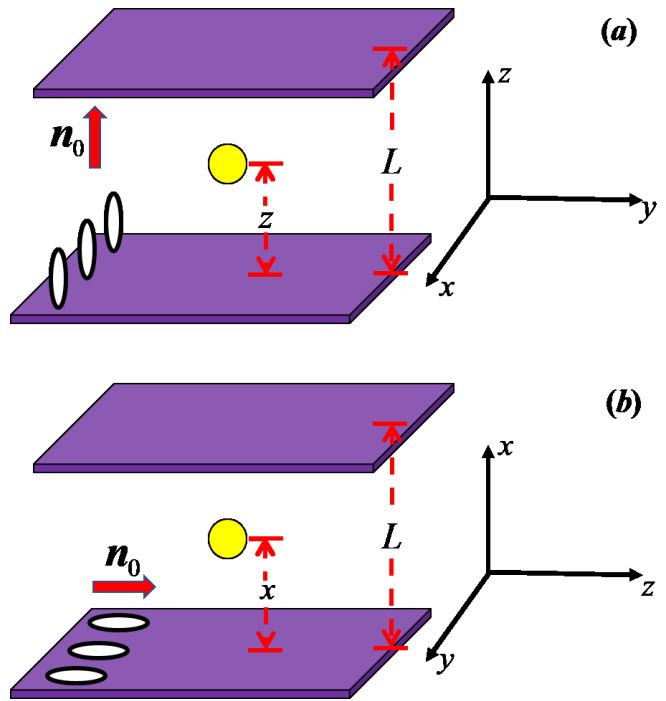}
  \caption{(Color online) Sketch of a microparticle suspended in a nematic cell with (a) homeotropic anchoring and (b) planar anchoring in the presence of an external field.\label{schematic}}
\end{figure}
The suspended microparticle induces a director distortion from the undeformed director field $\bf{n}_0$ = (0,0,1), expressed by a two-component deviation field $n_{\mu}$ ($\mu=x,y$), both small at regions far away from the microparticle. Here we deliberately use two different coordinate frames, as shown in Figs.~1(a) and (b), so that the same set of symbol subscripts ($n_{\mu}$ ($\mu=x,y$)) can be usable under approximations for the two surface anchoring conditions throughout this paper. Assuming $\bf{n}$ $\approx$ ($n_{x}$,$n_{y}$,1) with one Frank constant approximation, the effective elastic energy for the system reads~\cite{S.B.Chernyshuk2012}
\begin{align}
U_{e} &=  K \int d^3x \Bigl[\frac{(\nabla n_{\mu})^2}{2} - \frac{k^2}{2}(\textbf{e}\cdot \textbf{n})^2 - 4\pi P(\textbf{x})\partial_{\mu}n_{\mu} \notag\\
  & \qquad - 4\pi C(\textbf{x})\partial_z\partial_{\mu}n_{\mu} \Bigr],\label{eq:elastic}
\end{align}
where $K$ is the Frank constant, $n_{\mu}$ ($\mu=x,y$) represents the components of the director field $\textbf{n}$ perpendicular to $\textbf{n}_0$, $P(\textbf{x})$ and $C(\textbf{x})$ denote the dipole- and the quadrupole moment densities as functions of position $\textbf{x}$ respectively, and $k^2=(4\pi K)^{-1}\Delta \varepsilon E^2$ with $\Delta\varepsilon=\varepsilon_{\|}-\varepsilon_{\bot}$ the dielectric anisotropy of the NLC, which can be positive or negative with both cases considered in this paper. Here $\varepsilon_{\|}$ and $\varepsilon_{\bot}$ are the dielectric susceptibilities of the liquid crystal molecule parallel and perpendicular to the molecular long axis respectively. For homeotropic anchoring, when an electric field is applied along $z$ axis (Fig.~1(a)), the Euler-Lagrange equations are given by~\cite{S.B.Chernyshuk2012}
\begin{equation}\label{ELz}
\Delta n_{\mu} - k^2 n_{\mu} = 4\pi[\partial_{\mu} P(\textbf{x}) -\partial_z\partial_{\mu} C(\textbf{x})].
\end{equation}
On the other hand for planar (homogeneous) anchoring, as shown in Fig. 1(b), when an external electric field is applied parallel to $x$ axis, we have the Euler-Lagrange equations written as~\cite{S.B.Chernyshuk2012}
\begin{equation}\label{ELx}
\Delta n_{\mu} + k^2 \delta_{x\mu} n_{\mu} = 4\pi[\partial_{\mu} P(\textbf{x}) -\partial_z\partial_{\mu} C(\textbf{x})].
\end{equation}
If the applied electric field is parallel to $y$ axis (Fig. 1(b)), then the Euler-Lagrange equations are~\cite{S.B.Chernyshuk2012}
\begin{equation}\label{ELy}
\Delta n_{\mu} + k^2 \delta_{y\mu} n_{\mu} = 4\pi[\partial_{\mu} P(\textbf{x}) -\partial_z\partial_{\mu} C(\textbf{x})].
\end{equation}
With Dirichlet boundary conditions $n_{\mu} ({\bf{s}}) = 0$ on the two walls, the solution to Euler-Lagrange equations can be written as~\cite{S.B.Chernyshuk2012}
\begin{equation}\label{nu}
n_{\mu}(\textbf{x}) = \int_V d^3 \textbf{x}^{'} G_{\mu}{(\textbf{x},\textbf{x}^{'})} [-\partial_{\mu}^{'} P(\textbf{x}^{'}) + \partial_{\mu}^{'}\partial_{z}^{'}C(\textbf{x}^{'})],
\end{equation}
where $G_\mu$ is the Green's function for $n_\mu$. Please be noted that here $\mu$ in the integral does not follow Einstein summation notation.

\section{RESULTS AND DISCUSSIONS}
Given the Green's functions and the total energies in the Appendix for five different cases, as follows we can plot the energy profiles as a function of microparticle position for different electric fields. The occurrence of positional transition triggered by external electric field shall also be discussed.
\subsection{Homeotropic boundary condition}
\subsubsection{External field perpendicular to the two plates}
Now we first consider an electric field applied perpendicular to the two plates of the NLC cell with homeotropic anchoring condition, i.e., $\textbf{E}\Vert z$ in Fig.~1(a), and $\Delta\varepsilon>0$, a Fr\'{e}edericksz-like positional transition is found and the effects of cell thickness, Frank elastic constant and microparticle size on the threshold value triggering the positional transition have been discussed in our previous paper~\cite{K.Xiao2019}.
Furthermore, in order to study the effect of microparticle density on the critical electric value, we plot the threshold value against $\sqrt{K}/L$ for different microparticle densities (0.99, 1.0 and 1.03 $\rm g\cdot cm^{-3}$) in Fig.~2, where a Fr\'{e}edericksz curve (black) is shown as well. It is clearly seen that the critical electric field for positional transition for different microparticle densities shows a Fr\'{e}edericksz-like behavior, exhibiting straight lines nearly parallel to each other, yet with a different slope from the Fr\'{e}edericksz transition curve (black). A further calculation shows that such a Fr\'{e}edericksz-like linear master curve of critical electric field does not depend or negligibly depend on the density of the microparticle, and the proposed formula for the critical electric field in Ref.~\cite{K.Xiao2019} gives a prefactor of $3\sqrt{\pi}$ in comparison with the Fr\'{e}edericksz transition threshold expression. In the case when $\Delta\varepsilon<0$,
\begin{figure}[htp]
  \includegraphics[width=\linewidth,keepaspectratio]{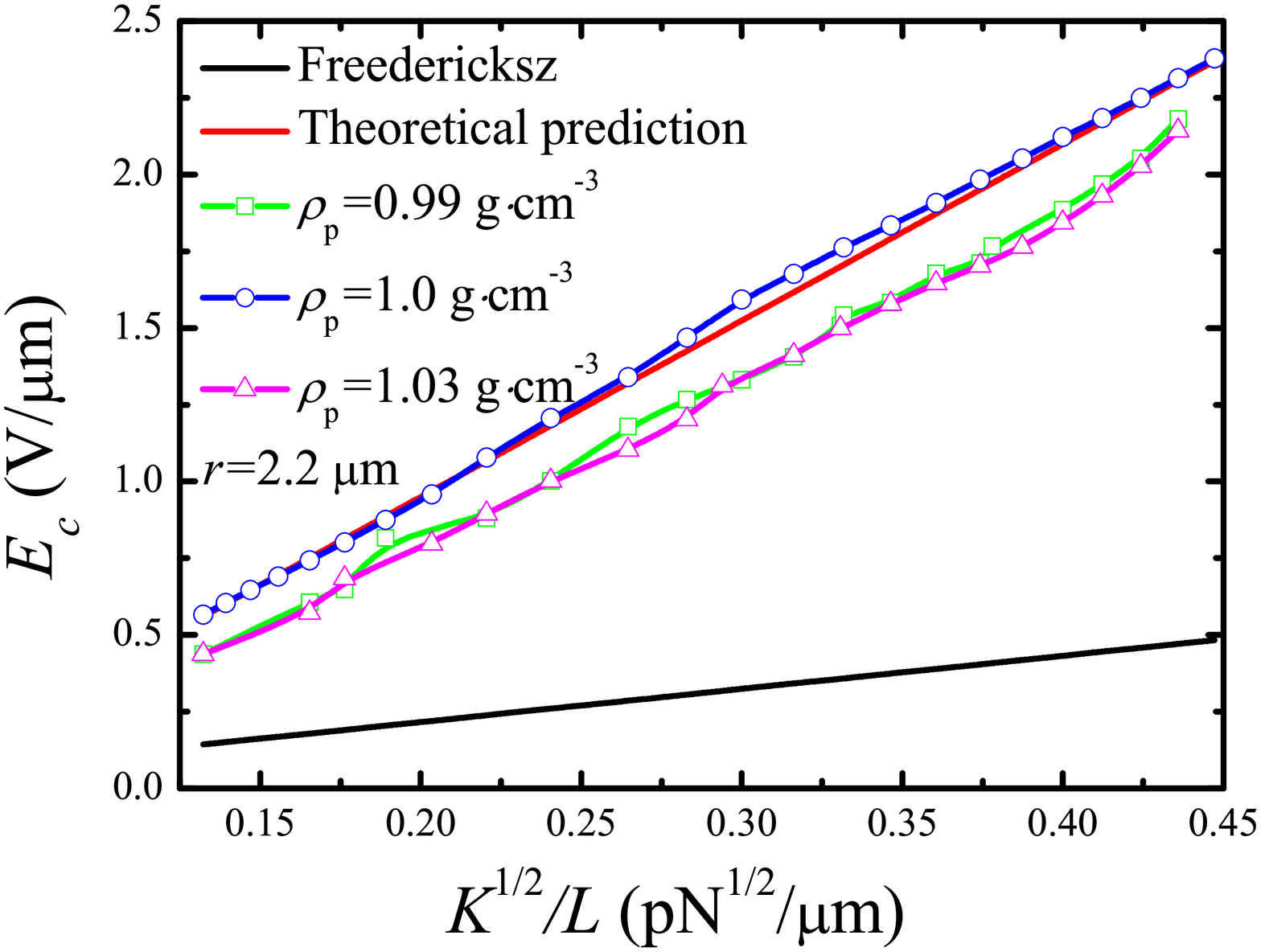}
  \caption{(Color online) Dependence of $E_c$ and $\sqrt{K}/L$ for different densities of microparticle (0.99, 1.0 and 1.03 $\rm g\cdot cm^{-3}$).}
\end{figure}
it is found that the microparticle is trapped at the midplane of the NLC cell, indicating that an application of external electric field does not trigger a positional phase transition. This is because when $\textbf{E}\Vert z$ and $\Delta\varepsilon<0$, the realignment of liquid crystal molecules with the increase of the electric field narrows down the interaction potential well rather than flatten it, which creates a force directing toward the midplane much larger than the gravitational contribution and thus denies any positional transition.
\subsubsection{External field parallel to the two plates}
In the case when the external field is parallel to the two plates, i.e., $\textbf{E}\Vert x$ in Fig.~1(a), surprisingly, the microparticle is trapped in the midplane of the NLC cell regardless of the sign of the molecular dielectric anisotropy, indicating that an application of external electric field, however large it is, can not trigger a positional transition. This can be understood by considering the fact that the molecular long (short) axes tend to align along the direction of applied electric field as $\Delta\varepsilon>0$ ($\Delta\varepsilon<0$). As we increase the field applied, the interaction potential is found to be narrowed down, corresponding to a strong midplane-directing restoring force. Therefore for the homeotropic boundary condition, the positional transition occurs only in an NLC cell with positive molecular dielectric anisotropy when the external electric field is applied along the undeformed director field.
\subsection{Planar boundary condition}
\subsubsection{External field perpendicular to the two plates}
When LC molecules are horizontally anchored on the two cell walls, as depicted in Fig. 1(b), let us first consider a positive dielectric anisotropy case $\Delta\varepsilon>0$ when an electric field is applied vertically
\begin{figure}[htp]
  \includegraphics[width=\linewidth,keepaspectratio]{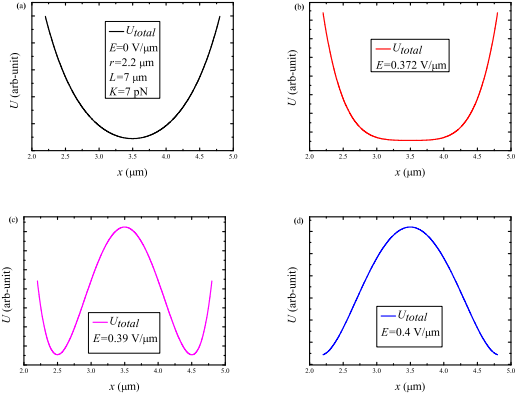}
  \caption{(Color online) Total energy profile as a function of the suspended microparticle position for an NLC cell with planar anchoring in the presence of different electric fields perpendicular to the two plates.}
\end{figure}
 to the two plates, i.e., along $x$ axis in the figure. Intriguingly, a significant feature is observed regarding the profile of total energy as a function of microparticle position for four different electric fields, as illustrated in Fig.~3. In the presence of small field (below the critical electric value), Figs.~3(a) and (b) show that the interaction potential well around the midplane tends to be flattened in this region due to the realignment of liquid crystal molecules made by the increment of external electric field.
However, when the electric field rises beyond the threshold value, there exists two symmetric equilibrium positions for the suspended microparticle (see Figs.~3(c) and (d)). Which one the microparticle shifts to is decided by the perturbation stemming from the asymmetric buoyant force, i.e. by the density difference between NLC and microparticle ($\rho_{LC}-\rho_{mp}$). Notably, the total energy now is almost equal to the elastic energy due to the fact that the gravitational contribution is much smaller in contrast to the elastic one, generating the depths of the two local minimums in Fig.~3(c) (and Fig.~3(d)) nearly equal to each other.
\begin{figure}
  \includegraphics[width=\linewidth]{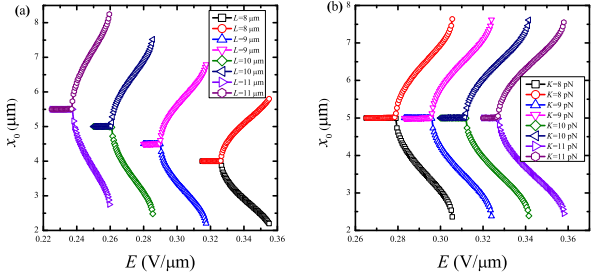}
  \caption{(Color online) Equilibrium position $x_0$ in response to electric field for different (a) cell thicknesses (8, 9, 10 and 11 $\rm \mu m$) with Frank elastic constant $K=7$ $\rm pN$ and the radius of microparticle $r=2.2$ $\rm \mu m$, and (b) Frank elastic constants (8, 9, 10 and 11 $\rm pN$) with cell thickness $L=10$ $\rm \mu m$ and radius of microparticle $r=2.2$ $\rm \mu m$.}
\end{figure}

To probe the influence of cell thickness and Frank constant on the critical field value, we plot the equilibrium position of the suspended microparticle against the applied electric field for different cell thicknesses (8, 9, 10 and 11 $\rm \mu m$) and Frank elastic constants (8, 9, 10 and 11 pN), as shown in Figs.~4(a) and (b), where a positional transition occurs at some electric field threshold values and there exist two bistable equilibrium positions when the external field applied exceeds the critical value. A more deeper investigation, as later shown in Fig.~(5), exhibits that the critical value of the external electric field is inversely proportional to $L$ and linearly proportional to $\sqrt{K}$, a Fr\'{e}edericksz-like behavior.

As a following step, we examine whether the critical electric value is correlated with the size and density of the microparticle.
Surprisingly,
Figs.~5(a) and (b) show that the plots of the equilibrium position of suspended microparticle against the applied electric field for different microparticle sizes and densities overlap each other, suggesting that the critical electric value is independent of or negligibly depends on microparticle size and density.
To gain more insight into the dynamic behaviors of the microparticle, we further plot the threshold value against $\sqrt{K}/L$ in Figs.~5(c) and (d), where a Fr\'{e}edericksz curve (black) is shown as well.
It is interesting to find that the critical electric field to trigger a positional transition for
\begin{figure}[htp]
  \includegraphics[width=\linewidth,keepaspectratio]{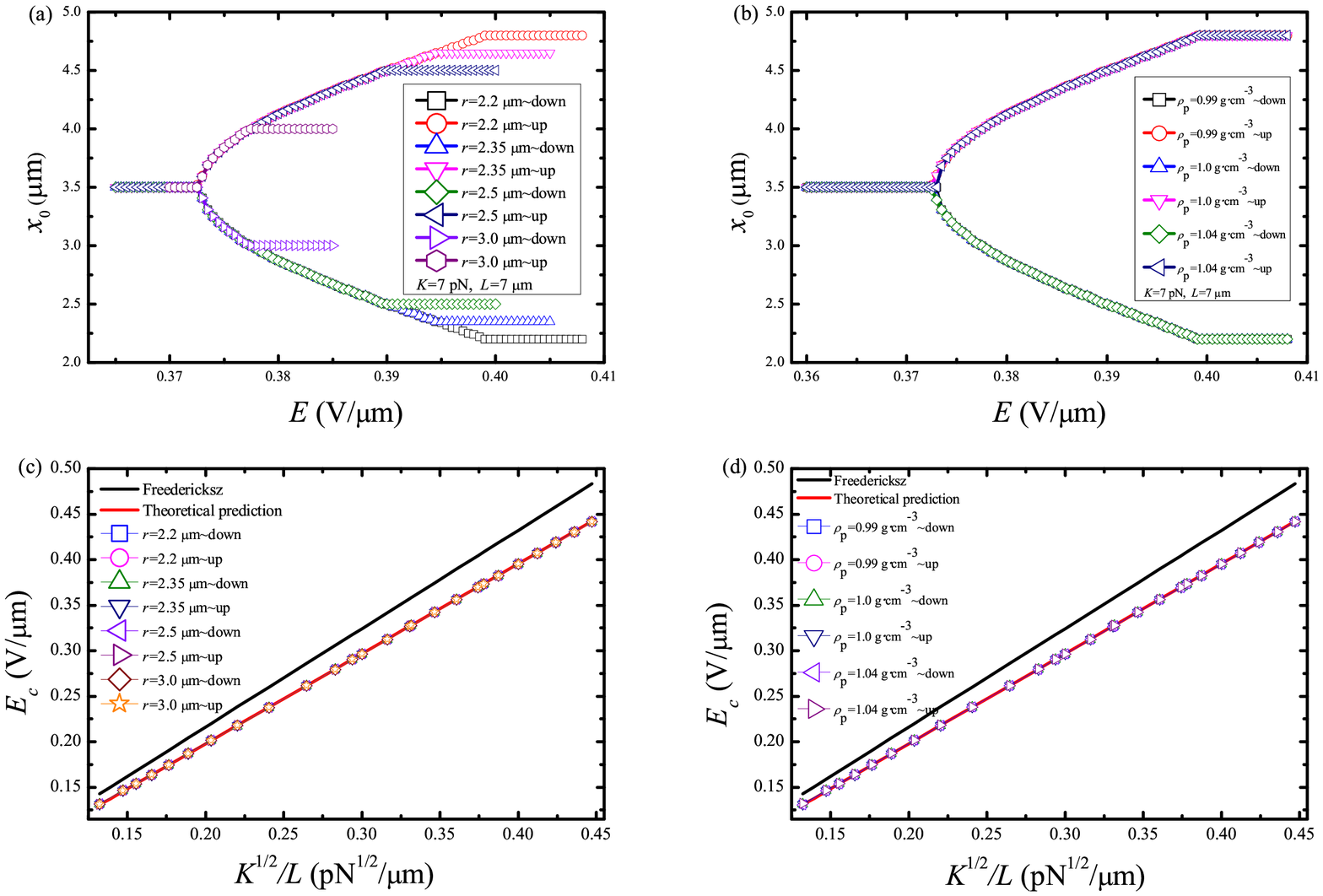}
  \caption{(Color online) Equilibrium position $x_0$ for different (a) radii (2.2 $\rm \mu m$, 2.35 $\rm \mu m$, 2.5 $\rm \mu m$ and 3.0 $\rm \mu m$); (b) densities (0.99, 1.0 and 1.04 $\rm g\cdot cm^{-3}$) of a microparticle with $K=7$ pN and $L=7$ $\rm \mu m$, showing the same critical value $E_c$ of electric field triggering positional transition. The dependence of $E_c$ on $\sqrt{K}/L$ for different (c) radii (2.2 $\rm \mu m$, 2.35 $\rm \mu m$, 2.5 $\rm \mu m$ and 3.0 $\rm \mu m$); (d) densities (0.99, 1.0 and 1.04 $\rm g\cdot cm^{-3}$) of the microparticle, obeying strictly a master curve which can be given by the theoretical prediction Eq.~\eqref{eq:prediction1}.}
\end{figure}
 a microparticle suspended in an NLC cell follows a Fr\'{e}edericksz-like linear master curve with slightly different slopes, a universal one also valid for different microparticle sizes and densities.

By comparing the numerical calculation results with the Fr\'{e}edericksz transition ($\pi \sqrt{4\pi/|\Delta\varepsilon|}\sqrt{K}/L$) in Figs.~5(c) and (d),  we found that the slope difference between them is by a prefactor of $\sim$0.915, and that enables us to propose a theoretical prediction for the critical electric field
\begin{equation}\label{eq:prediction1}
E_{c} \simeq 0.915{\mathcal{F}}\\,
\end{equation}
where $\mathcal{F}$ denotes the Fr\'{e}edericksz\ effect. Such a prediction, as shown by straight line (red) in Figs.~5(c) and (d), agrees very well for different radii (2.2, 2.35, 2.5, and 3.0 $\rm \mu m$) and densities (0.99, 1.0 and 1.04 $\rm g\cdot cm^{-3}$) of microparticle. Due to the mathematical difficulty, we still don't know how to derive 0.915 analytically.

In the case when $\Delta\varepsilon<0$, it is found that the suspended microparticle is trapped at the midplane of the NLC cell, which can be predicted by the profile change of the total energy potential well due to application of an external electric field in the vertical direction ($x$ direction in Fig.~1(b)). The short axes of liquid crystal molecules tend to align along the electric field, a result leading to the narrowing of total potential well and thereby generating strong restoring force acting on the suspended microparticle. Thus, in the case of a microparticle suspended in an NLC cell with planar anchoring condition in the presence of an external electric field applied perpendicular to the two plates, the positional transition triggered by the electric field may occur only under the condition of positive molecular dielectric anisotropy.

\subsubsection{External field parallel to the two plates but perpendicular to the anchoring direction}
Now let us consider the case when the electric field is applied parallel to the two plates but perpendicular to the anchoring direction, i.e., along $y$ axis in Fig.~1(b). It is found that no matter $\Delta\varepsilon>0$ or $\Delta\varepsilon<0$, the microparticle is always trapped at the midplane of the NLC cell regardless
\begin{figure}[htp]
  \includegraphics[width=\linewidth,keepaspectratio]{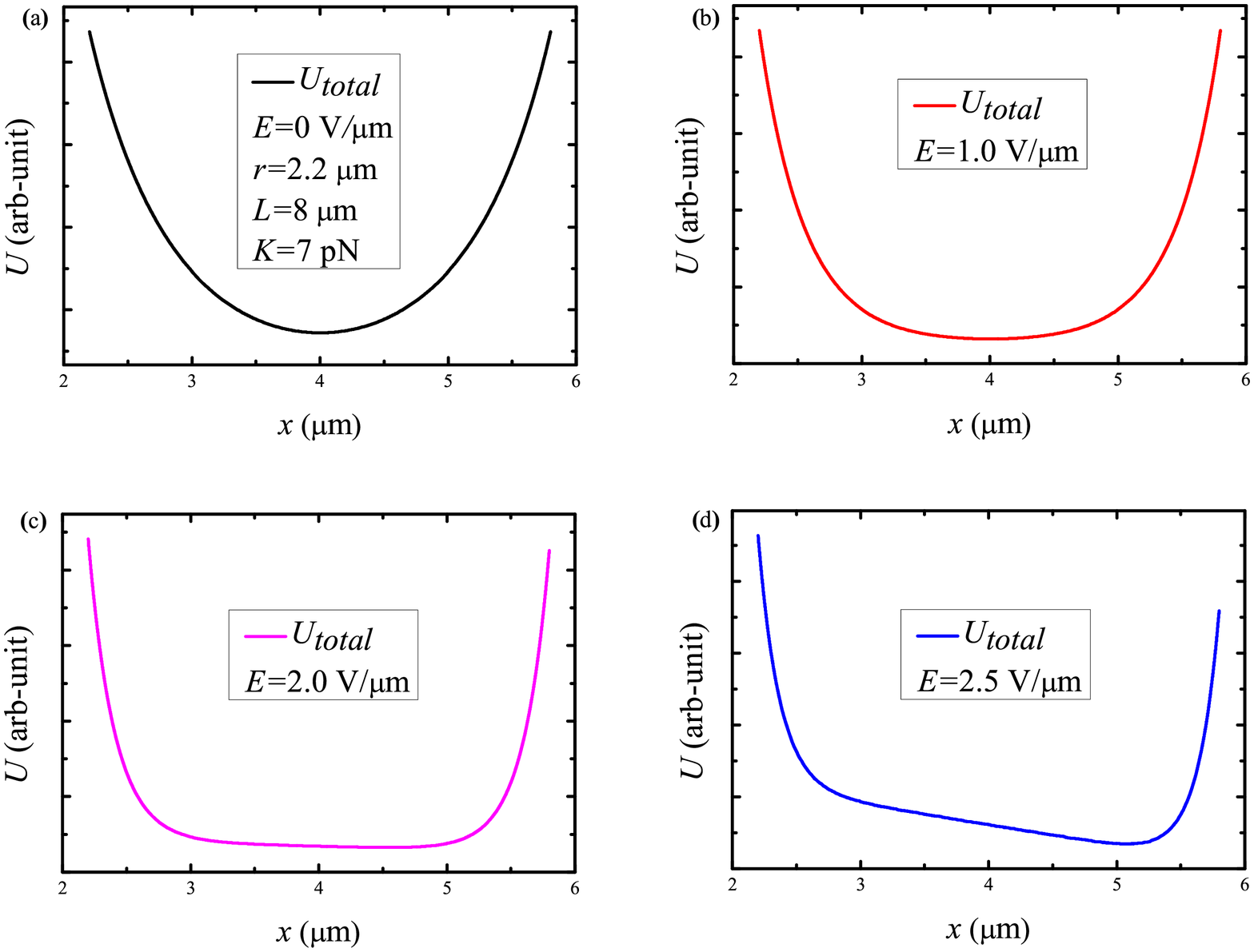}
  \caption{(Color online)  Total energy profile as a function of the suspended microparticle position for an NLC cell with planar anchoring in the presence of four chosen electric fields parallel to the two plates and the anchoring direction as well.}
\end{figure}
 of the magnitude of the electric field applied, indicating that no positional transition occurs. The reason lies in that the realignment of the liquid crystal molecules in the presence of the external electric field does not flatten the interaction potential well substantially enough so as to decrease its corresponding equivalent restoring force on the microparticle to a small magnitude, with which the asymmetric gravitational force becomes competitive.
\subsubsection{External field parallel to the two plates and the anchoring direction}
Finally, we consider an NLC cell in the presence of an electric field parallel to the two plates and the anchoring direction as well, i.e., $\bf{E}$$\|$$z$ in Fig.~1(b). If given a positive molecular dielectric anisotropy, namely $\Delta\varepsilon>0$, we can plot, as shown in Fig. 6, the total energy profile as a function of the suspended microparticle position for four chosen electric fields. In the presence of a small external field, the total energy profile remains symmetric, indicating that the elastic interaction among LC molecules dominates the LC alignment, especially in the region close to the midplane. Thus the contribution made by asymmetric gravitational potential is trivial if compared with elasticity and the suspended microparticle will be trapped within its midplane, as demonstrated in Figs.~6(a) and (b).
While as the electric field is increased, it is found that it tends to widen and flatten the bottom of the elastic potential well, which equivalently by contrast amplifies the relative contribution made by the asymmetric buoyant force to the total energy of the NLC cell. As a result, the buoyant force will drive the microparticle with ease from the midplane to a new equilibrium position (see Figs.~6(c) and (d)). It is apparent that the sign of $\rho_{LC}-\rho_{mp}$ determines the direction of the microparticle displacement. It looks very much like that the bottom of the interaction potential well around the midplane is "pressed" due to the realignment of liquid crystal molecules made by the applied external field, which creates a "fast lane" along the vertical direction in the cell for the suspended microparticle to migrate. Once such a "fast lane" constructed by the external field in the cell reaches a critical value of "smoothness" (corresponding to a weakened elastic energy gradient), driven by the asymmetric buoyant force, it triggers a positional transition for the suspended microparticle from the midplane to its new equilibrium position.

In order to study the influence of cell thickness and Frank constant on the critical value of electric field, plots for the equilibrium position for the suspended microparticle against the applied electric field for different cell thicknesses (8, 10, 12 and 15 $\mu m$) and Frank elastic constants (8, 10, 12 and 15 pN) are presented in Figs.~7(a) and (b),
\begin{figure}[htp]
  \includegraphics[width=\linewidth,keepaspectratio]{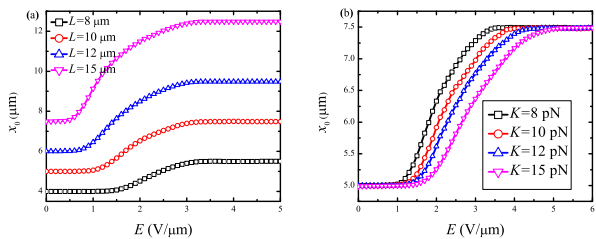}
  \caption{(Color online) Equilibrium position $x_0$ in response to electric field for different (a) cell thicknesses (8, 10, 12 and 15 $\rm \mu m$), where the Frank elastic constant and the radius of microparticle are set as $K=7$ pN and $r=2.5$ $\rm \mu m$, and (b) Frank elastic constants (8, 10, 12 and 15 pN), where the cell thickness and the radius of microparticle are set as $L=10$ $\rm \mu m$ and $r=2.5$ $\rm \mu m$.}
\end{figure}
where it is found that a positional transition occurs when the external field applied exceeds a threshold value. It is also shown that the thinner the cell thickness $L$ is and the larger the Frank elastic constant $K$ is, the larger the critical electric field is needed to trigger the positional transition.

In a similar way to the previous sections, the dependence of the threshold value on microparticles size and density is also investigated.
\begin{figure}[htp]
  \includegraphics[width=\linewidth,keepaspectratio]{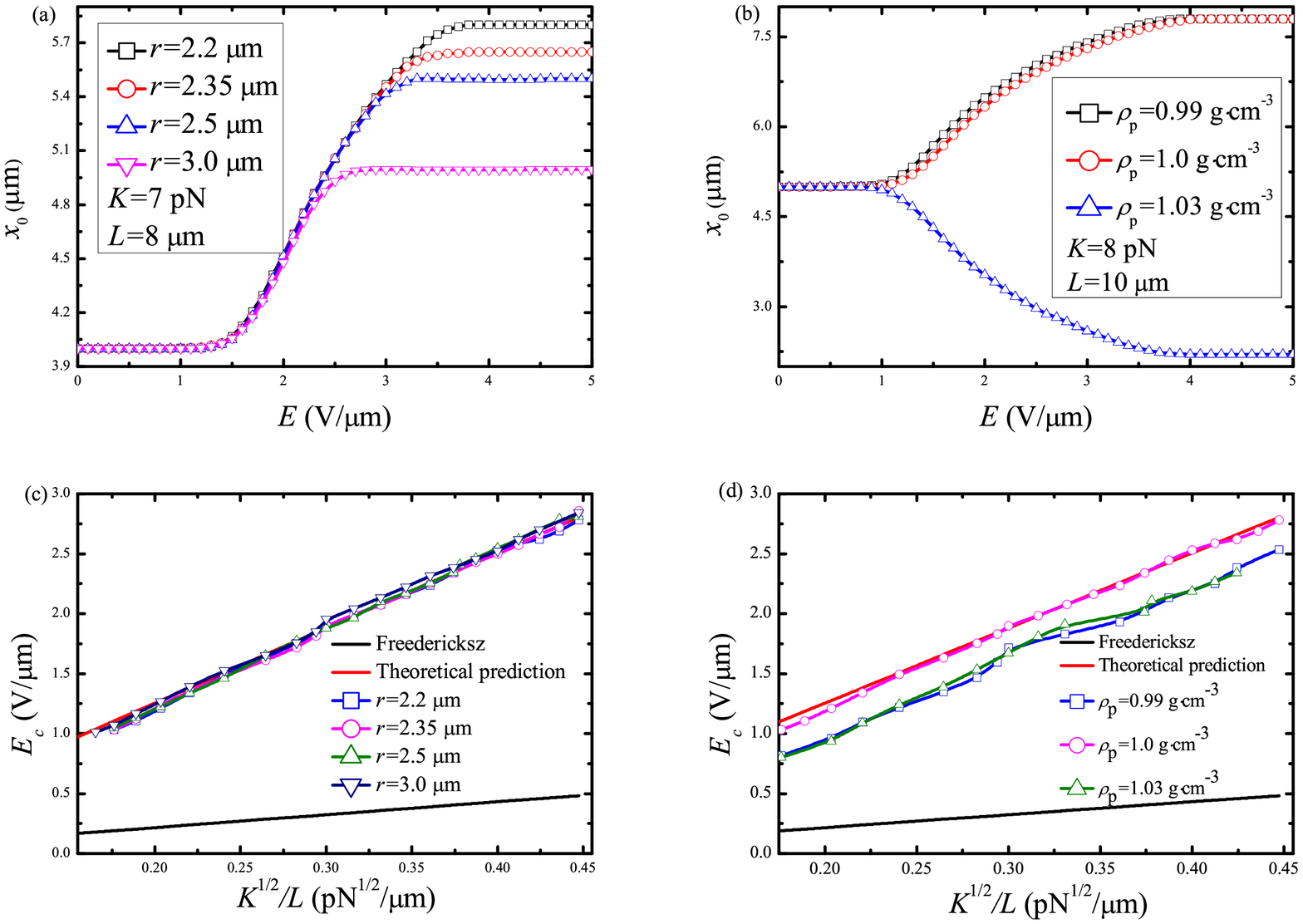}
  \caption{(Color online) (a) Equilibrium position $x_0$ for different radii of microparticle with $K=7$ pN and $L=8$ $\rm \mu m$, showing the same critical value $E_c$ of electric field triggering positional transition. (b) $K=8$ pN and $L=10$ $\rm \mu m$. The dependence of $E_c$ and $\sqrt{K}/L$ for different (c) radii (2.2 $\rm \mu m$, 2.35 $\rm \mu m$, 2.5 $\rm \mu m$ and 3.0 $\rm \mu m$); (d) densities (0.99, 1.0 and 1.03 $\rm g\cdot cm^{-3}$) of microparticle, obeying strictly a master curve given by theoretical prediction Eq.~\eqref{eq:prediction2}.}
\end{figure}
Figs.~8(a) and (b) depict the equilibrium position against the applied electric field for different microparticle sizes and densities, where the overlapping of equilibrium position in Fig.~8(a) suggests that the critical electric value is almost independent of microparticle size. Whereas the symmetry of the equilibrium position of microparticle with density equal to 0.99 $\rm g\cdot cm^{-3}$ and 1.03 $\rm g\cdot cm^{-3}$ in Fig.~8(b) indicates that the slope of the master curve of critical electric value is nearly independent of the magnitude of equivalent microparticle density. To gain more insight into the dynamic behaviors of the microparticle, the threshold value is plotted against $\sqrt{K}/L$ in Figs.~8(c) and (d), where a Fr\'{e}edericksz transition curve (black) is shown as well. The existence of slightly difference instead of overlapping to each other for the equilibrium position of microparticle with density equal to 0.99 $\rm g\cdot cm^{-3}$ and 1.0 $\rm g\cdot cm^{-3}$ in Fig.~8(b), leads to different intercepts of the Fr\'{e}edericksz-like linear master curves for critical electric field in Fig. 8(d). The further study of electric field threshold shows that it seems to be inversely proportional to cell thickness $L$ and proportional to the root square of Frank elastic constant $K$, a behavior similar to the field threshold of Fr\'{e}edericksz phase transition. Like before, the critical electric field for a positional transition to occur for a microparticle suspended in a NLC cell remains unchanged for different microparticle sizes and densities.

Similarly, a contrast between the numerical calculation results and the traditional Fr\'{e}edericksz transition curve($\pi \sqrt{4\pi/|\Delta\varepsilon|}\sqrt{K}/L$) in Figs.~8(c) and (d) shows that the slope difference between them is by a prefactor of $\sim 5.8$. More specifically, an explicit expression
\begin{equation}\label{eq:prediction2}
E_{c} \simeq 5.8{\mathcal{F}}-0.08\\
      = 5.8\pi\sqrt{\frac{4\pi K}{|\Delta\varepsilon| L^2}}-0.08
\end{equation}
for critical electric field can be proposed as a theoretical prediction. Such a prediction, as shown by straight line (red) in Figs.~8(c) and (d), agrees very well for different radii (2.2, 2.35, 2.5, and 3.0 $\rm \mu m$) and densities (0.99, 1.0 and 1.03 $\rm g\cdot cm^{-3}$) of microparticle. This once again verifies the conclusion that the critical electric field is independent of microparticle size, of which the reason might lie in that in the present theoretical model, the microparticle is approximately treated as a dipole in the far field expansion.

As for the case $\Delta\varepsilon<0$ when the external field applied parallel to both the two plates and the anchoring direction, i.e., $\textbf{E}\Vert z$ in Fig.~1(b), a bistable equilibrium state structure is found as the electric field exceeds a threshold value, as illustrated in Fig.~9.
\begin{figure}[htp]
  \includegraphics[width=\linewidth,keepaspectratio]{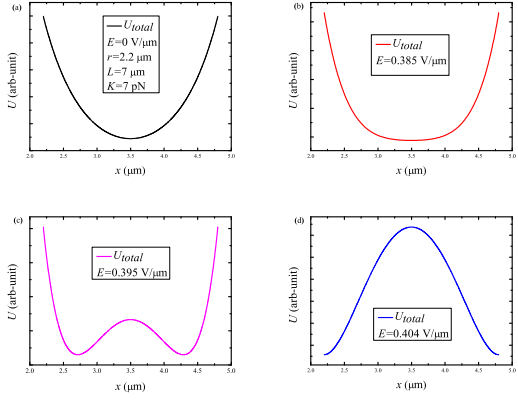}
  \caption{(Color online) Total energy profile as a function of the suspended microparticle position for different external electric fields.}
\end{figure}
In the small-field region, the external field applied tends to, first of all, flatten the bottom of potential well, as shown in Figs.~9(a) and (b). Further increase of external field will change the one-state potential structure to a bistable one. As the gravitational contribution to the total energy is still negligibly small compared to the elastic one, one sees no involvement of gravitational force to the determination of the critical value of positional transition for the microparticle in the NLC cell. Thus, the positional transition in this case does not come from the competition between the gravitational force and the equivalent elastic force but rather purely from the bistable local minimum of the elastic potential, as shown in Figs.~9(c) and (d). Nevertheless the asymmetric gravitational force still plays a very important role in determining the direction of microparticle motion (up or down) by acting as a small but significant perturbation, or more precisely, by the sign of buoyant force (the sign of $\rho_{LC}-\rho_{mp}$). Therefore, the magnitude of the asymmetric gravitational force in this case is trivial but not its sign.

 In order to understand how cell thickness and Frank elastic constant affect the critical value of electric field, we plot equilibrium position against the applied electric field for different cell thicknesses (7, 8, 9 and 10 $\rm \mu m$) and Frank elastic constants (8, 9, 10 and 11 pN), as shown in Figs.~10(a) and (b),
\begin{figure}[htp]
  \includegraphics[width=\linewidth,keepaspectratio]{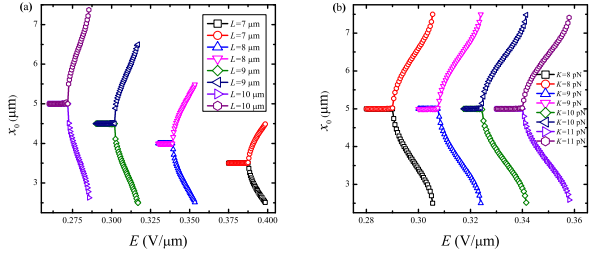}
  \caption{(Color online) Equilibrium position $x_0$ in response to electric field for different (a) cell thicknesses (7, 8, 9 and 10 $\mu m$), where the Frank elastic constant and the radius of micro-droplet are set as $K=7$ pN and $r=2.5$ $\mu m$, and (b) Frank elastic constants (8, 9, 10 and 11 pN), where the cell thickness and the radius of micro-droplet are set as $L=10$ $\mu m$ and $r=2.5$ $\mu m$.}
\end{figure}
 where a bifurcation of equilibrium position is found due to the bistable state structure of elastic potential and a positional transition occurs when the external field applied reaches a threshold value.

Finally, in order to gain more insights into the physics hidden behind the dynamic behaviors of microparticle, it is worthwhile to evaluate whether the critical electric value is correlated with the size and density of the microparticle.
The dependence of the equilibrium position on the applied electric field for different microparticle sizes and densities is shown in Figs.~11(a) and (b),
\begin{figure}[htp]
  \includegraphics[width=\linewidth,keepaspectratio]{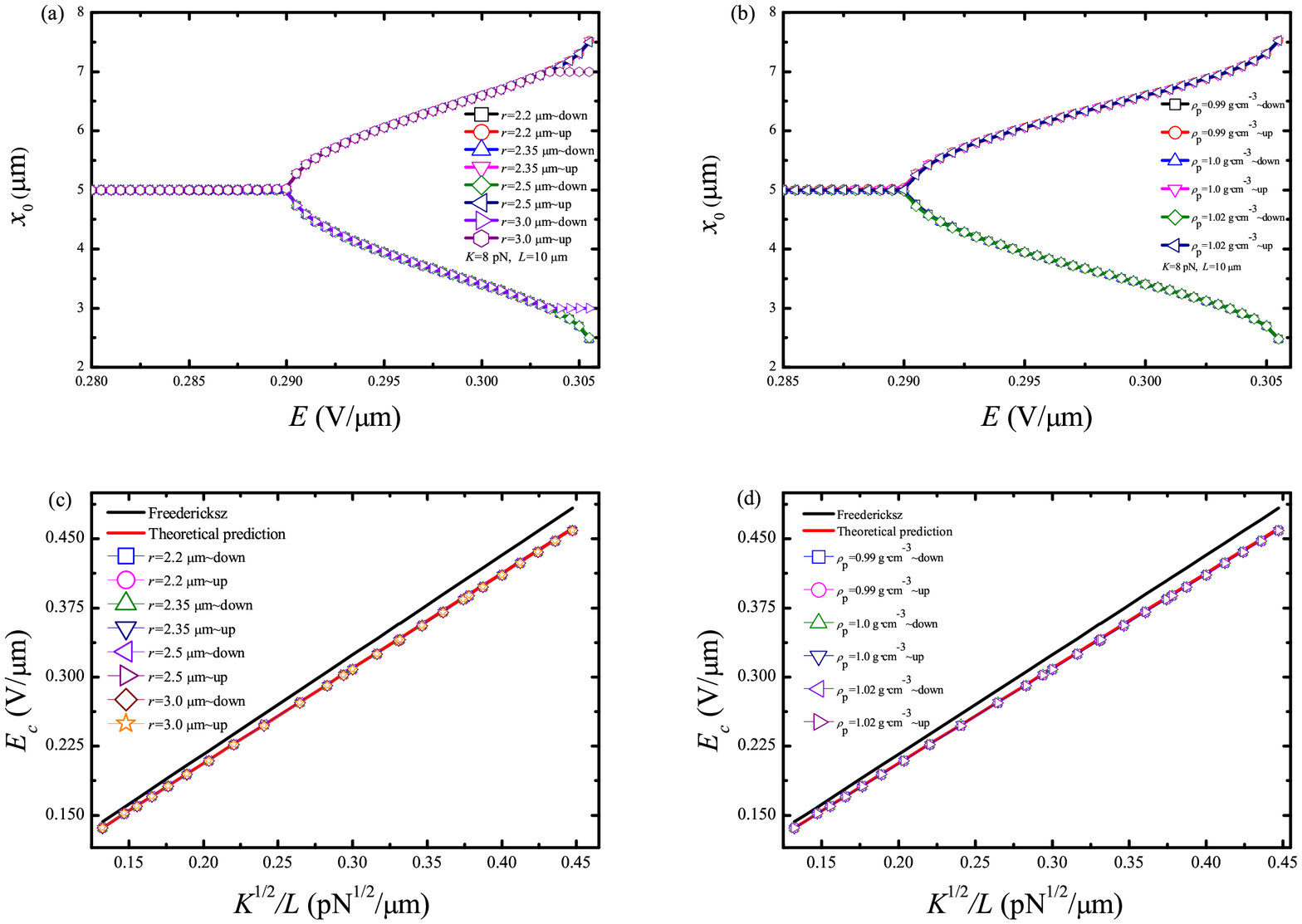}
  \caption{(Color online) Equilibrium position $x_0$ for different (a) radii (2.2 $\rm \mu m$, 2.35 $\rm \mu m$, 2.5 $\rm \mu m$ and 3.0 $\rm \mu m$); (b) densities (0.99, 1.0 and 1.02 $\rm g\cdot cm^{-3}$) of microparticle with $K=8$ pN and $L=10$ $\rm \mu m$, showing the same critical value $E_c$ of electric field triggering positional transition. The dependence of $E_c$ and $\sqrt{K}/L$ for different (c) radii (2.2 $\rm \mu m$, 2.35 $\rm \mu m$, 2.5 $\rm \mu m$ and 3.0 $\rm \mu m$); (d) densities (0.99, 1.0 and 1.02 $\rm g\cdot cm^{-3}$) of microparticle, obeying strictly a master curve given by theoretical prediction Eq.~\eqref{eq:prediction3}.}
\end{figure}
where the strict overlapping of equilibrium position in the figures implies that the critical electric value is, as shown in the previous section, independent of microparticle size and density.
For a better understanding of the dynamic behaviors of the microparticle, we further plot the threshold value against $\sqrt{K}/L$ in Figs.~11(c) and (d), with a Fr\'{e}edericksz transition curve (black) shown as well. It is found that the critical electric field triggering a positional transition for a microparticle suspended in a NLC cell follows a Fr\'{e}edericksz  master curve irrelevant to microparticle size and density.

More precisely, by comparing the numerical calculation results with the Fr\'{e}edericksz effect curve ($\pi \sqrt{4\pi/|\Delta\varepsilon|}\sqrt{K}/L$) in Fig.~11(c) and (d), it is found that the slope difference between them is by a prefactor of $\sim 3/{\pi}$, leading to a proposed theoretical prediction
\begin{equation}\label{eq:prediction3}
E_{c} \simeq \frac{3}{{\pi}}{\mathcal{F}}\\
      = 6\sqrt{\frac{\pi K}{|\Delta\varepsilon| L^2}}
\end{equation}
for the critical electric field.
\begin{table*}[htbp]

\caption{Formation of a vertical fast "lane" for positional transition to occur ($+$) and not to occur ($-$) for a microparticle suspended in an NLC cell in the presence of an external electric field.}
\centering
\renewcommand\arraystretch{1.5}
\begin{tabular}{@{}ccp{60pt}<{\centering}p{80pt}<{\centering}p{60pt}<{\centering}@{}}
\toprule
\toprule
\multirow{2}{*}{Anchoring}   & \multirow{2}{*}{\makecell[c]{Molecular dielectric \\ anisotropy}} & \multicolumn{3}{c}{Field direction}                                                             \\ \cmidrule(l){3-5}
                             &                                        & $\textbf{E}$ $\perp$ Plates & \multicolumn{2}{c}{$\textbf{E}$ $\|$ Plates}                                  \\ \midrule
\multirow{2}{*}{Homeotropic} & $\Delta\varepsilon>0$                       & +                      & \multicolumn{2}{c}{$-$}                                                  \\
                             & $\Delta\varepsilon<0$                          & $-$                      & \multicolumn{2}{c}{$-$}                                                  \\
\multirow{3}{*}{Planar}      &                                        &                        & $\textbf{E}$ $\perp$ Anchoring & \multicolumn{1}{l}{$\textbf{E}$ $\|$ Anchoring} \\
                             & $\Delta\varepsilon>0$                       & +/bistable             & $-$                          & +                                         \\
                             & $\Delta\varepsilon<0$                          & $-$                      & $-$                          & +/bistable                                \\ \bottomrule
                             \bottomrule
\end{tabular}
\end{table*}
Such a prediction, as shown by straight line in Figs.~11(c) and (d), agrees very well for different radii (2.2, 2.35, 2.5, and 3.0 $\rm \mu m$) and densities (0.99, 1.0 and 1.02 $\rm g\cdot cm^{-3}$) of microparticle.

Based on the discussions in the sections above, it is quite obvious that the external electric field applied enhances the existing anisotropy of distortion generated by the boundaries of the NLC cell shaped by the the movable suspended microparticle and the two parallel walls. It looks like there exists an anisotropic movable "bubble" surrounding the suspended microparticle, created by the external field and the boundary conditions combined. Inside the "bubble" along the vertical direction a fast "lane" will be constructed once the external field applied reaches a critical value. The electric field threshold is a signal to complete the construction and a "key" to switching on the use of the fast "lane", wobbling the "bubble" along the vertical direction, and thereby tune the motion of the microparticle inside, which has been proved to be a positional transition~\cite{K.Xiao2019}. Interestingly, this kind of motion can be found in the some SciFi novels picturing one of the possible tactics for intergalactic travel in the future by moving a planetary object via wobbling the space-time around it, which is supported by general relativity. After a thorough discussion of all the conditions combined to create such a wobbling "bubble" in a NLC cell in the presence of an external electric field, we come up with a table for a positional transition to occur in such a system, as shown in Table~I. It is found in the table that out of the ten combinations of field direction, molecular dielectric anisotropy, and anchoring feature, only four shows the possible occurrence of positional transition. Moreover, for a nematic liquid crystal cell with planar surface alignment, a bistable equilibrium structure for the transition is found when the direction of applied electric field is (a) perpendicular to the two plates of the cell with positive molecular dielectric anisotropy, or (b) parallel to both the two plates and the anchoring direction of the cell with negative molecular dielectric anisotropy.
\section{CONCLUSION}
In summary, using the Green's function method, the total energy for a microparticle suspended in an NLC cell in the presence of an external electric field is calculated. It is found that with the application of the external electric field, it is possible to create an anisotropic bubble around the microparticle with a vertical fast "lane" for the microparticle to move from the midplane to a new equilibrium position. Such a new equilibrium position is decided via a competition between the buoyant force and the effective force built upon the microparticle inside the "lane". The threshold value of external field, which triggers positional transition under appropriate conditions of surface anchoring feature, field direction and molecular dielectric anisotropy, depends on thickness $L$ and Frank elastic constant $K$ and slightly on the microparticle size and density, in a Fr\'{e}edericksz-like manner as coined by the authors before, but by a factor. For an NLC cell with planar surface alignment, a bistable equilibrium structure for the transition is found when the direction of the applied electric field is (a) perpendicular to the cell wall with positive molecular dielectric anisotropy, and (b) parallel to the undeformed director field $\bf{n}_0$ of the NLC cell with negative molecular dielectric anisotropy. Except for the positional transition, when the electric field applied is parallel to the two plates and perpendicular to the anchoring direction, the microparticle suspended in NLC will be trapped in the midplane, regardless of the sign of the molecular dielectric anisotropy. Explicit formulae proposed for the critical electric field agrees extremely well with the numerical calculation.

\begin{acknowledgements}
This work was funded by the National Science Foundation of China under Grant No. 11974292, No.11974291, and No. 11947401.
\end{acknowledgements}

\section{Appendix: Green's functions and total energies}
\subsection{Homeotropic boundary condition}
Here we first consider an NLC cell sandwiched between two parallel plates, where LC molecules are homeotropically anchored and the coordinate $z$ axis is chosen along the normal direction of the two plates, as shown in Fig. 1(a).
\subsubsection{External field perpendicular to the two plates}
In this case, when an electric field is applied perpendicular to the two plates, i.e., $\bf{E}$$\|$$z$ in Fig. 1(a), the corresponding Euler-Lagrange equations are written as Eq.~\eqref{ELz}. With Dirichlet boundary conditions $n_{\mu}{(z=0)} = n_{\mu}{(z=L)} = 0$, the Green's function can be derived as~\cite{S.B.Chernyshuk2012}
\begin{align}
G_{\mu}{(\textbf{x},\textbf{x}^{'})} =& \frac{4}{L} \sum_{n=1}^{\infty}\sum_{m=-\infty}^{\infty} e^{im(\varphi - {\varphi}^{'})} \sin\frac{n\pi z}{L}\notag\\
 & \times\sin\frac{n\pi z^{'}}{L} I_{m}(\lambda_{n}\rho_{<})K_{m}(\lambda_{n}\rho_{>}),\label{eq:GFI}
\end{align}
where $\varphi$ and ${\varphi}^{'}$ are the azimuthal angles, $z$ and $z^{'}$ are the positional coordinates, $I_{m}$ and $K_{m}$ are modified Bessel functions, $\rho_{<}$ is the smaller one between $\sqrt{x^2+y^2}$ and $\sqrt{x^{'2}+y^{'2}}$, and $\lambda_{n}=[({n\pi}/{L})^2+{\Delta \varepsilon E^2}/{4\pi K}]^{1/2}$ with $L$ the thickness of the NLC cell.
Using the definition of self energy given in terms of Green's function~\cite{S.B.Chernyshuk2012}
\begin{align}
U_{dd}^{self} = -2\pi Kp^{2}\partial_{\mu}\partial_{\mu}^{'}H_{\mu}(\textbf{x},\textbf{x}^{'})|_{\textbf{x}=\textbf{x}^{'}},\label{eq:Uself}
\end{align}
 where $H_{\mu}{(\textbf{x},\textbf{x}^{'})} = G_{\mu}{(\textbf{x},\textbf{x}^{'})} - {1}/{|\textbf{x} - \textbf{x}^{'}|}$, we obtain the elastic energy $U_{e}^{I}$ for a microparticle suspended in an NLC cell in the presence of an electric field. Besides the elastic energy, the gravitational potential $U_g$ due to buoyant force should be considered as well, leading to a total energy written as
\begin{align}
U_{total}^I =& U_{e}^I+U_{g}\notag\\
=& -2\pi K p^2\left[-\frac{4}{L}\sum_{n=1}^\infty \lambda_{n}^2 \sin^2(\frac{n\pi z}{L})K_{0}(\lambda_n \rho)+\frac{1}{\rho^3}\right]_{\rho\rightarrow 0}\notag\\
&-\frac{4}{3}\pi r^3 (\rho_{LC}-\rho_{mp})gz,\label{eq:Utotal}
\end{align}
where $r$ is the radius of microparticle, $p=2.04r^2$ is the magnitude of the equivalent dipole moment, $\rho_{LC}-\rho_{mp}$ is the density difference between liquid crystal and microparticle, and $g=9.8$ $\rm m/s^2$ is the gravitational acceleration.
\subsubsection{External field parallel to the two plates}
For the case of an electric field parallel to the two plates, i.e., $\bf{E}$$\|$$x$ in Fig. 1(a), the Euler-Lagrange equations for $n_x$ and $n_y$ are written as Eq.~\eqref{ELx}. With Dirichlet boundary conditions $n_{\mu}{(z=0)} = n_{\mu}{(z=L)} = 0$, the related Green's functions $G_x$ and $G_y$ are given by
\begin{align}
G_{x}{(\textbf{x},\textbf{x}^{'})} =& \frac{4}{L} \sum_{n=1}^{\infty}\sum_{m=-\infty}^{\infty} e^{im(\varphi - {\varphi}^{'})} \sin\frac{n\pi z}{L}\notag\\
 & \times\sin\frac{n\pi z^{'}}{L} I_{m}(\nu_{n}\rho_{<})K_{m}(\nu_{n}\rho_{>}),\notag\\
G_{y}{(\textbf{x},\textbf{x}^{'})} =& \frac{4}{L} \sum_{n=1}^{\infty}\sum_{m=-\infty}^{\infty} e^{im(\varphi - {\varphi}^{'})} \sin\frac{n\pi z}{L}\notag\\
 & \times\sin\frac{n\pi z^{'}}{L} I_{m}(\mu_{n}\rho_{<})K_{m}(\mu_{n}\rho_{>}),\label{eq:GFII}
\end{align}
 where $\nu_{n}=[({n\pi}/{L})^2-{\Delta \varepsilon E^2}/{4\pi K}]^{1/2}$ and $\mu_{n}=n\pi/L$. In analogue to the previous case, we can obtain the elastic energy $U_{e}^{II}$ and the total energy $U_{total}^{II}$ is written as
\begin{align}
U_{total}^{II} =& U_{e}^{II}+U_{g}\notag\\
=& -2\pi K p^2\biggl[-\frac{2}{L}\sum_{n=1}^\infty \sin^2(\frac{n\pi z}{L})(\alpha_n + \beta_n)\notag\\
&+\frac{1}{\rho^3}\biggr]_{\rho\rightarrow 0}-\frac{4}{3}\pi r^3 (\rho_{LC}-\rho_{mp})gz,\label{eq:UtotalII}
\end{align}
where $\alpha_n = \nu_n^2 K_{0}(\nu_n \rho) + \mu_n^2 K_{0}(\mu_n \rho)$, and $\beta_n = \nu_n^2 K_{2}(\nu_n \rho) - \mu_n^2 K_{2}(\mu_n \rho)$.

\subsection{Planar boundary condition}
\subsubsection{External field perpendicular to the two plates}
When an electric field is applied vertically to the two plates, i.e., $\bf{E}$$\|$$x$ in Fig.~1(b), the Euler-Lagrange equations are given by Eq.~\eqref{ELx}, and the corresponding Green's functions $G_x$ and $G_y$ read as~\cite{S.B.Chernyshuk2012}
\begin{align}
G_{x}{(\textbf{x},\textbf{x}^{'})} =& \frac{4}{L} \sum_{n=1}^{\infty}\sum_{m=-\infty}^{\infty} e^{im(\varphi - {\varphi}^{'})} \sin\frac{n\pi x}{L}\notag\\
 & \times\sin\frac{n\pi x^{'}}{L} I_{m}(\nu_{n}\rho_{<})K_{m}(\nu_{n}\rho_{>}),\notag\\
G_{y}{(\textbf{x},\textbf{x}^{'})} =& \frac{4}{L} \sum_{n=1}^{\infty}\sum_{m=-\infty}^{\infty} e^{im(\varphi - {\varphi}^{'})} \sin\frac{n\pi x}{L}\notag\\
 & \times\sin\frac{n\pi x^{'}}{L} I_{m}(\mu_{n}\rho_{<})K_{m}(\mu_{n}\rho_{>}), \label{eq:GFIII}
\end{align}
 with $\nu_{n}$ and $\mu_{n}$ identical to those in Eq.~\eqref{eq:GFII}. Similarly, the elastic energy $U_e^{III}$ can be obtained and the total energy $U_{total}^{III}$ can be derived as
\begin{align}
U_{total}^{III} =& U_{e}^{III}+U_{g}\notag\\
=& -2\pi K p^2 \biggl[ \frac{4}{L}\sum_{n=1}^\infty \mu_{n}^2 \big[\cos^2(\frac{n\pi x}{L})K_{0}(\nu_{n}\rho)-\frac{1}{2}\notag\\
&  \times \sin^2(\frac{n\pi x}{L})\big(K_{0}(\mu_{n}\rho) - K_{2}(\mu_{n}\rho) \big)\big]+\frac{1}{\rho^3} \biggr]_{\rho\rightarrow 0} \notag\\
& - \frac{4}{3}\pi r^3 (\rho_{LC}-\rho_{mp})gx,\label{eq:UtotalIII}
\end{align}
where $x$ denotes the vertical position of the microparticle.
\subsubsection{External field parallel to the two plates but perpendicular to the anchoring direction}
When the electric field applied is parallel to the two plates but perpendicular to the anchoring direction, i.e., $\bf{E}$$\|$$y$ in Fig.~1(b), the Euler-Lagrange equations can be given by Eq.~\eqref{ELy}, with their corresponding Green's functions $G_x$ and $G_y$ written as
\begin{align}
G_{x}{(\textbf{x},\textbf{x}^{'})} =& \frac{4}{L} \sum_{n=1}^{\infty}\sum_{m=-\infty}^{\infty} e^{im(\varphi - {\varphi}^{'})} \sin\frac{n\pi x}{L}\notag\\
 & \times\sin\frac{n\pi x^{'}}{L} I_{m}(\mu_{n}\rho_{<})K_{m}(\mu_{n}\rho_{>}),\notag\\
G_{y}{(\textbf{x},\textbf{x}^{'})} =& \frac{4}{L} \sum_{n=1}^{\infty}\sum_{m=-\infty}^{\infty} e^{im(\varphi - {\varphi}^{'})} \sin\frac{n\pi x}{L}\notag\\
 & \times\sin\frac{n\pi x^{'}}{L} I_{m}(\nu_{n}\rho_{<})K_{m}(\nu_{n}\rho_{>}),\label{eq:GFIV}
\end{align}
with the same $\nu_{n}$ and $\mu_{n}$ as those in Eq.~\eqref{eq:GFIII}. In a similar way, the total energy $U_{total}^{IV}$ is given by
\begin{align}
U_{total}^{IV} =& U_{e}^{IV}+U_{g}\notag\\
=& -2\pi K p^2 \biggl[ \frac{4}{L}\sum_{n=1}^\infty \mu_{n}^2 \cos^2(\frac{n\pi x}{L})K_{0}(\mu_{n}\rho)\notag\\
& -\frac{2}{L}\sum_{n=1}^\infty \nu_{n}^2 \sin^2(\frac{n\pi x}{L}) \big(K_{0}(\nu_{n}\rho)-K_{2}(\nu_{n}\rho)\big)\notag\\
& +\frac{1}{\rho^3} \biggr]_{\rho\rightarrow 0} - \frac{4}{3}\pi r^3 (\rho_{LC}-\rho_{mp})gx,\label{eq:UtotalIV}
\end{align}
where $U_{e}^{IV}$ is the elastic energy.
\subsubsection{External field parallel to the two plates and the anchoring direction}
Finally, if an electric field is applied parallel to the two plates and the planar anchoring direction as well, i.e., $\bf{E}$$\|$$z$ in Fig.~1(b). Given the corresponding Euler-Lagrange equations Eq.~\eqref{ELz}, the Green's functions are~\cite{S.B.Chernyshuk2012}
\begin{align}
G_{\mu}{(\textbf{x},\textbf{x}^{'})} =& \frac{4}{L} \sum_{n=1}^{\infty}\sum_{m=-\infty}^{\infty} e^{im(\varphi - {\varphi}^{'})} \sin\frac{n\pi x}{L}\notag\\
 & \times\sin\frac{n\pi x^{'}}{L} I_{m}(\lambda_{n}\rho_{<})K_{m}(\lambda_{n}\rho_{>}),\label{eq:GFV}
\end{align}
with the same $\lambda_{n}$ as that in  Eq.~\eqref{eq:GFI}. Similarly, the total energy $U_{total}^{V}$ can be derived as
\begin{align}
U_{total}^{V} =& U_{e}^{V}+U_{g}\notag\\
=& -2\pi K p^2 \biggl[ \frac{4}{L}\sum_{n=1}^\infty (\frac{n\pi}{L})^2 \cos^2(\frac{n\pi x}{L})K_{0}(\lambda_{n}\rho)\notag\\
& -\frac{2}{L}\sum_{n=1}^\infty \lambda_{n}^2 \sin^2(\frac{n\pi x}{L}) \big(K_{0}(\lambda_{n}\rho)-K_{2}(\lambda_{n}\rho)\big)\notag\\
& +\frac{1}{\rho^3} \biggr]_{\rho\rightarrow 0} - \frac{4}{3}\pi r^3 (\rho_{LC}-\rho_{mp})gx,\label{eq:UtotalV}
\end{align}
with $U_{e}^{V}$ the elastic energy.

\end{document}